\documentclass{article}
\usepackage{spconf,amsmath,graphicx}
\usepackage{hyperref}
\usepackage{booktabs}  
\usepackage{adjustbox}
\usepackage{caption}  
\usepackage{xcolor}
\usepackage{multirow}

\newcommand{\etal}{\textit{et al.}}

\usepackage{amssymb}
\usepackage{pifont}
\newcommand{\cmark}{\ding{51}}%
\newcommand{\xmark}{\ding{55}}%
\newcommand{\greencheck}{{\color{green}\cmark}}
\newcommand{\redcross}{{\color{red}\xmark}}

\title{Improving Lip-synchrony in Direct\\Audio-Visual Speech-to-Speech Translation}
\name{
\begin{tabular}{@{}c@{}}
Lucas Goncalves, 
Prashant Mathur\sthanks{Corresponding authors.},
Xing Niu$^*$,
Brady Houston,
Chandrashekhar Lavania, \\
Srikanth Vishnubhotla\sthanks{Provided administrative and resourcing support.},
Lijia Sun$^\ddagger$,
Anthony Ferritto$^\ddagger$
\end{tabular}
}
\address{Amazon}

\begin{document}
\ninept
\maketitle
\begin{abstract}


\textit{Audio-Visual Speech-to-Speech Translation} (AVS2S) typically prioritizes improving translation quality and naturalness. However, an equally critical aspect in audio-visual content is lip-synchrony—ensuring that the movements of the lips match the spoken content—essential for maintaining realism in dubbed videos. Despite its importance, the inclusion of lip-synchrony constraints in AVS2S models has been largely overlooked. This study addresses this gap by integrating a lip-synchrony loss into the training process of AVS2S models. Our proposed method significantly enhances lip-synchrony in direct audio-visual speech-to-speech translation, achieving an average LSE-D score of 10.67, representing a 9.2\% reduction in LSE-D over a strong baseline across four language pairs. Additionally, it maintains the naturalness and high quality of the translated speech when overlaid onto the original video, without any degradation in translation quality.

\end{abstract}
\section{Introduction}
\label{sec:intro}

Traditionally, research has emphasized the importance of lip synchrony -- the alignment of translated audio with the visible mouth movements of the original actors—as a key factor in maintaining the quality and realism of dubbed content~\cite{chaume2012audiovisual, hu2021neural, kim2019neural}. However, improving lip synchrony should not compromise translation quality and naturalness \cite{perego2016empirical, brannon2022dubbing}. Simply achieving a trade-off between lip synchrony and translation quality may not significantly enhance the overall user experience. In this study, we aim to improve lip-synchrony in dubbed videos while ensuring that translated speech is natural and of high quality when overlaid on original videos, addressing the trade-off.

Recent advancements in audio-visual speech translation \cite{choi2024av2av} have focused on generating translated audio-visual outputs from audio-visual inputs. These approaches typically generate visuals by modifying lips to match the audio~\cite{prajwal2020lip} which can generate artifacts and lead to two main problems 1) generation of deepfake~\cite{datta2024exposinglipsyncingdeepfakesmouth} which raises ethical concern~\cite{deepfakedetectionchallenge} 2) generation of deepfake videos without safeguarding people's identities and personalities ('likeness') from being digitally recreated without their consent \cite{bariach2024faces}. Modifying a speaker's lip movements in automatic translations could violate these concerns, potentially infringing on their image rights \cite{murphy2023face, meskys2020regulating}. Moreover, generating high-quality video outputs in audio-visual speech translation currently poses significant challenges. State-of-the-art video generation models often produce artifacts, such as issues with teeth appearance \cite{xu2024vasa}, which can distract viewers and negatively affect the overall viewing experience. In this work, we sidestep the problem of modifying videos and our project solely focuses on improving speech translation while maintaining the original visual content. This approach avoid introducing any visual artifacts and maintain the integrity of the original video.

While early works incorporated visual information to improve translation quality of speech to speech translation models, none have actively worked maintaining lip synchrony in the generated target speech and original video \cite{huang2023avtranspeech, cheng2023mixspeech}. Current works like~\cite{choi2024av2av} do not explicitly utilize lip-synchrony as a constraint in model training, and the synchrony evaluations are performed only on the outputs where both audio-visual are generated. In this work, we explore the idea of leveraging visual inputs to enhance lip synchrony between generated speech in target language and original video, at the same time, side-stepping generation of visuals along with the speech. Our contribution focuses on integrating a lip-synchrony loss into training process of audio-visual (AV) speech translation models. As discussed in~\cite{brannon2022dubbing}, dubbers typically violate lip-sync in order to achieve better translation quality and naturalness in speech, so we explore ways to improve lip-synchrony in the training process without compromising on translation quality or naturalness of the translated speech.

\section{Related Work}
\label{sec:related_work}


Lip synchronization has emerged as a crucial research area with wide-ranging applications, particularly in automatic dubbing for translation \cite{chaume2012audiovisual}. Numerous studies have focused on improving synchrony in dubbing through various approaches, such as isometric translation, where translations are generated to match the length of the source text \cite{lakew2021isometricmt}, and prosodic alignment, which seeks to synchronize translated text with the source speech \cite{effendi2022,virkar2022prosodicalignmentoffscreenautomatic}. Additionally, joint training of translation and duration models has been explored to enable a single model to learn both translation and phone duration estimation \cite{pal2023improvingisochronousmachinetranslation,chronopoulou2023jointlyoptimizingtranslationsspeech}. While these methods effectively replicate the prosody of the source speech (speech-pause structure), they often overlook the critical aspect of lip synchronization in dubbing.

Prajwal et al. \cite{27_prajwal} introduced an approach to incorporate a lip-sync discriminator, resulting in more accurate synchronization between arbitrary video and audio inputs. However, the aforementioned approaches suffer from issues such as mouth blurring and inconsistent rendering of teeth. Other approaches, instead of directly manipulating lip movements based on audio, Xie et al. \cite{28_xie} proposed a two-stage framework. In the first stage, a generator is trained to predict facial landmarks from the audio. In the second stage, these predicted landmarks, combined with the target frame, are used to generate the final output. However, this method has its limitations: the generated reference landmarks are often inaccurate, and the approach may distort the actor's identity, similar to the method proposed by Choi et al. \cite{choi2024av2av}, where facial attributes are modified to match the generated audio content. One issue with these techniques is that there are ethical concerns about safeguarding individuals' identities and likenesses from being digitally recreated or modified without consent \cite{bariach2024faces}. Unlike the previous approaches, we explore a research area with two realistic constraints 1) where the original videos are preserved 2) voice characteristics of original speakers are not mimicked.




\begin{figure}[!t]
    \centering
    \includegraphics[width=0.48\textwidth]{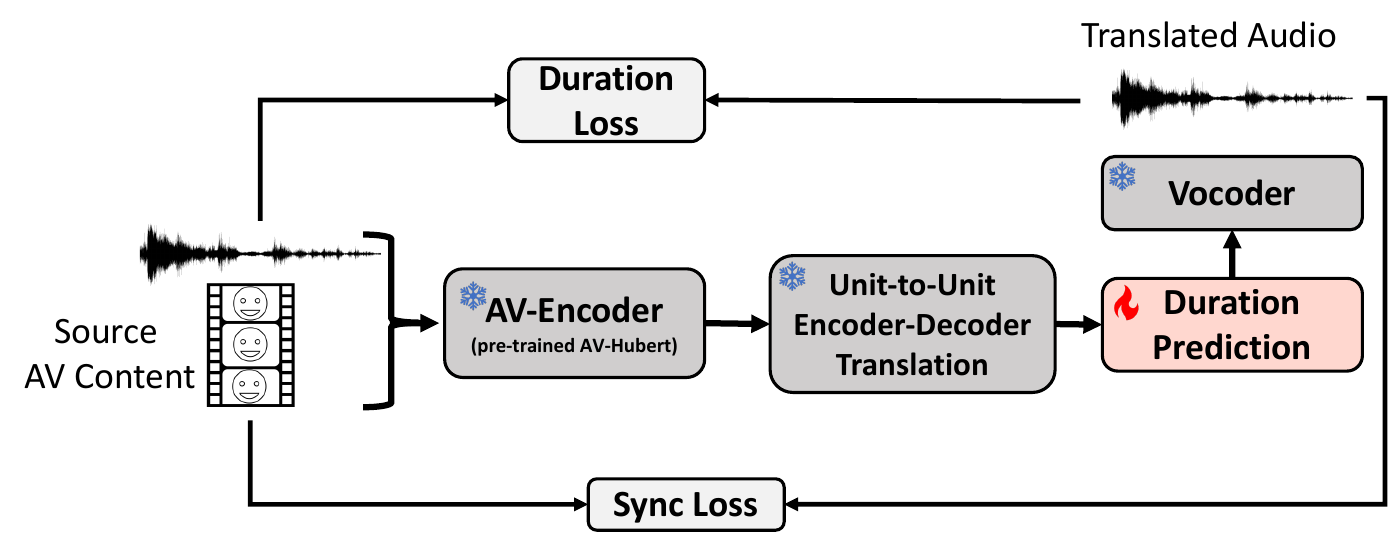}
    \caption{AVS2S Framework Overview}
    \label{fig:Model}
\end{figure}

\section{Methodology}
\label{sec:method}

Our overall framework of \emph{Audio-Visual Speech-to-Speech Translation} (AVS2S) system in depicted in Fig. \ref{fig:Model} and is based off Choi \etal ~\cite{choi2024av2av}. The framework consists of the following: visuals and speech content from the original video are fed to an Audio-Visual (AV) Encoder which processes lip region and speech content and convert them into discrete unified audio-visual units. AV-Encoder used in this work is a pre-trained multilingual AV-HuBERT \cite{shi2021learning}, presented in the work of Choi \etal \cite{choi2024av2av}. Next, we have a translation module which is a encoder-decoder network that translates source language AV unit to target language AV unit based on the work of Kim \etal \cite{kim2023many}. Lastly, the vocoder is based on unit-based HiFiGAN framework \cite{kong2020hifi, lee2021direct}. Essentially, we remove the visual generation component from the AV Renderer \cite{choi2024av2av} and overlay the generated target language speech on the original video.

In Choi \etal \cite{choi2024av2av}, AV Renderer component generates both speech and visuals at the same time as they take the AV unit as input and feed it to vocoder and wav2lip~\cite{prajwal2020lip} modules, which results in an inherent lip-synchrony as both modules are utilizing same AV unit as input. In our case, since we are not synthesizing face and instead overlaying the generated speech on original video, the same lip-synchrony element is lost.  In this work, we aim to enhance lip-synchrony between the translated output audio stream of an input video (with faces) from one language to another using an AVS2S framework. The following section provides details about the AVS2S framework which adds two specific losses for lip-synchronization. 


\subsection{Duration Predictor}
\label{ssec:dur_pred}

Since the unit-to-unit translation framework takes input as source AV units and outputs target AV units, a duration predictor is typically employed before the vocoder which pre-processes the units needed to generate the target speech content. This predictor estimates the duration of each speech unit and adjusts them accordingly based on the desired target duration \cite{lee2021direct,kim2023many,choi2024av2av}. This is done via a ``length predictor" in AV2AV work, however, this model does not take into account duration of source language speech. In our initial experiments, we saw AV2AV model generating videos shorter than source and this problem can be attributed to length predictor not using source speech duration. In our work, since the output acoustic stream must be synchronized with the source video, we use a standard duration loss, computed between the source audio and the generated target speech, as defined in Eq. \ref{eq:dur}.



\begin{equation}
\mathcal{L}_{\text{dur}} = \frac{1}{N} \sum_{i=1}^{N} \left(\log d^{p}_i - \log d_i\right)^2
\label{eq:dur}
\end{equation}

\noindent where \( N \) is the total number of speech units in the sequence, \( d \) is the target duration, and \( d^{p}\) is the predicted duration.


In addition to duration loss, we also employ a synchronization loss, computed using the SyncNet model \cite{chung2016out} as our AV sync expert. Similar to the approach used in Wav2Lip \cite{prajwal2020lip}, the synchronization loss is defined as shown in Eq. \ref{eq:sync}.

\begin{equation}
\mathcal{L}_{\text{sync}} = \frac{1}{N} \sum_{i=1}^{N} \log \left(\text{SyncNet}^i(AV)\right)
\label{eq:sync}
\end{equation}

\noindent where \( \text{SyncNet}^i(AV) \) represents the synchronization score for the \(i\)-th AV pair (source video and generated audio).

The overall loss used to fine-tune the duration predictor depicted on the orange block in Fig. \ref{fig:Model} is then formulated as:

\begin{equation}
\mathcal{L}_{\text{total}} = \mathcal{L}_{\text{sync}} + \lambda \cdot \mathcal{L}_{\text{dur}}
\label{eq:total}
\end{equation}

\section{Experimental Settings}
\label{sec:experiments}

\begin{table}[t]
\centering
\begin{adjustbox}{width=0.8\columnwidth,center}
\begin{tabular}{l|c|c|c}
\textbf{Set} & \textbf{\#Videos} & \textbf{Total Duration} & \textbf{Avg Duration} \\ \hline
Trainval & 4,004  & 30 hr & 3.42 sec \\ \hline
Test & 412 & 51 min & 2.32 sec \\
\end{tabular}
\end{adjustbox}
\caption{LRS3 dataset statistics.}
\label{tab:lrs3_stats}
\end{table}

\subsection{Dataset}
\label{ssec:dataset}

We leverage LRS3 \cite{afouras2018deep} which is a large-scale video data consisting of thousands of spoken sentences collected from TED talks. The dataset statistics are provided in Table \ref{tab:lrs3_stats}. Importantly, there is no overlap between the videos used in the test set and those used in the Trainval sets. In our experiments, we use the Trainval set to fine-tune the duration predictor, and perform evaluation on the test set.

\begin{table*}[h!]
\centering
\begin{adjustbox}{max width=\textwidth}
\renewcommand{\arraystretch}{1.5} 
\setlength{\tabcolsep}{6pt} 
\begin{tabular}{l|c||c|c|c||c|c|c||c|c|c||c|c|c}
 &  & \multicolumn{3}{c||}{\textbf{En-Es}} & \multicolumn{3}{c||}{\textbf{En-Pt}} & \multicolumn{3}{c||}{\textbf{En-It}} & \multicolumn{3}{c}{\textbf{En-Fr}} \\ \hline
 & \textbf{Source} & \textbf{Synthetic} & \textbf{AV2AV} & \textbf{Ours} & \textbf{Synthetic} & \textbf{AV2AV}  & \textbf{Ours} & \textbf{Synthetic} & \textbf{AV2AV} &  \textbf{Ours} & \textbf{Synthetic} & \textbf{AV2AV} &  \textbf{Ours} \\ \hline
\textbf{LSE-C $\uparrow$} & 7.63 & 2.22 & 2.13 & \textbf{2.45*} & 2.27 & 1.15 & \textbf{3.43*} & 2.24 & 2.23 & \textbf{2.97*} & 2.08 & 2.23 & \textbf{2.46}  \\ \hline
\textbf{LSE-D $\downarrow$} & 6.88 & 12.08 & 11.6 & \textbf{10.68*} & 12.12 & 12 & \textbf{10.12*} & 11.92 & 11.67 & \textbf{10.89*} & 11.97 & 11.74 & \textbf{10.98*}  \\
\end{tabular}
\end{adjustbox}
\caption{Lip-synchrony evaluations of our proposed method (Ours) against baseline models (Synthetic and AV2AV-Speech) across four language pairs (English to Spanish (Es), Portuguese (Pt), Italian (It) and French (Fr).}
\label{tab:results}
\end{table*}

\subsection{Evaluation Metrics}
\label{ssec:metrics}

We evaluate our system using several key metrics to ensure high-quality audio-visual lip-synchronization, accurate speech translation and audio quality (naturalness). The LSE-C (Lip Sync Error - Confidence)~\cite{prajwal2020lip} metric measures the average confidence score, indicating the correlation between audio and video, with higher scores reflecting better AV synchronization. The LSE-D (Lip Sync Error - Distance) \cite{prajwal2020lip} metric calculates the distance between lip and audio representations, where lower scores signify better lip-synchronization. The Perceptual Evaluation of Speech Quality (PESQ) is an industry-standard metric for assessing audio quality, evaluating factors such as audio sharpness, call volume, background noise, clipping, and interference, with scores ranging from -0.5 to 4.5, where higher scores indicate better quality. This metric is essentially our proxy to measuring naturalness of speech. BLASER 2.0 \cite{barrault2023seamless} is a reference-free metric that evaluates end-to-end speech-to-speech translation by leveraging a multilingual multimodal encoder. It computes scores based on the similarity of input and output speech embeddings, which correlate well with human judgment. Finally, ASR-BLEU is a metric that measures BLEU score~\cite{papineni2002bleu} (translation quality) by comparing ASR outputs of generated speech to translations from Amazon Translate.\footnote{\url{https://aws.amazon.com/translate/}}

\subsection{Implementation Details}
\label{ssec:implement}

For visual feature pre-processing, we follow a similar approach to previous works by cropping the mouth region using a face detector and a facial landmark detector \cite{choi2024av2av, shi2021learning}. The audio is sampled at 16kHz. In training, as shown in Fig. \ref{fig:Model}, we extract units for both inputs and targets using the frozen AV-Encoder, which can accept either audio-visual or audio/video-only inputs. 

We fine-tune the duration predictor, as depicted in Fig. \ref{fig:Model} within the orange box. At this stage, the unit-to-unit decoder is frozen, and the duration predictor is trained using the loss function from Eq. \ref{eq:total}, with $\lambda$ set to 10. The model is fine-tuned from the pretrained weights from \cite{choi2024av2av}, and we fine-tune duration predictor module for 200K iterations. To ensure accurate synchrony evaluation, the vocoder generates the entire audio output at each step. We process one sample at a time and accumulate the gradient to reach a batch size of 32. The model is optimized using AdamW~\cite{loshchilov2019decoupledweightdecayregularization} with a learning rate of 2E-4.

\section{Experimental Results}
\label{sec:e_results}

\subsection{Baselines}
\label{ssec:results}
We use the latest work of AV2AV~\cite{choi2024av2av} as a strong baseline for this work. This is the only open-source AV translation model that has research permissive license. In our experiments with AV2AV, we did not evaluate their full audio-visual outputs. Instead, we use the generated speech by AV renderer and overlay it on the original video, similar to our approach. This was done to ensure fairness to our approach and to avoid making any facial modifications to the original videos. We call this system as \textit{AV2AV-Speech}.

As another baseline (\textit{Synthetic}), we leveraged machine translations via Amazon Translate and generated speech via Amazon Polly\footnote{\url{https://aws.amazon.com/polly/}}, overlayed it on the original video to generate dubbed video. We make sure that the speech generated by Polly is of similar duration as that of source via Speech Markers feature in Polly.

\begin{table}[t]
\centering
\begin{adjustbox}{width=0.9\columnwidth,center}
\begin{tabular}{l|l|c|c|c}
\textbf{Language}                    & \textbf{System} & \textbf{PESQ}  & \textbf{BLASER} & \textbf{ASR-BLEU} \\ \hline
\multirow{2}{*}{\textbf{En-Es}}    & \textbf{AV2AV}  & 1.047 & 0.798  & 35.13    \\ \cline{2-5} 
                            & \textbf{Ours}   & 1.051 & 0.797  & 35.18    \\ \hline
\multirow{2}{*}{\textbf{En-Pt}} & \textbf{AV2AV}  & 1.043 & 0.709  & 28.17    \\ \cline{2-5} 
                            & \textbf{Ours}   & 1.050 & 0.708  & 28.53    \\ \hline
\multirow{2}{*}{\textbf{En-It}}    & \textbf{AV2AV}  & 1.037 & 0.732  & 25.42    \\ \cline{2-5} 
                            & \textbf{Ours}   & 1.052 & 0.732  & 25.71    \\ \hline
\multirow{2}{*}{\textbf{En-Fr}}     & \textbf{AV2AV}  & 1.041 & 0.767  & 21.34    \\ \cline{2-5} 
                            & \textbf{Ours}   & 1.044 & 0.764  & 21.34    \\
\end{tabular}
\end{adjustbox}
\caption{Speech naturalness and translation quality evaluations of our proposed method (Ours) against AV2AV-Speech baseline across four  language pairs (English to Spanish (Es), Portuguese (Pt), Italian (It) and French (Fr).}
\label{tab:donoharmresults}
\end{table}

\subsection{Results}
\label{ssec:results_disc}

The results presented in Table \ref{tab:results} provide lip-synchrony scores of our proposed method (\textit{Ours}) against two baseline approaches across multiple language pairs (English to Spanish, Portuguese, Italian, and French). The ``Source" column contains LSE-C and LSE-D scores for the original video which serves as a reference point for the best achievable lip-synchrony alignment (i.e. upper bound) between the generated speech and the original video. 

From the results in Table \ref{tab:results}, we see that the \textit{Synthetic} generation of speech, when overlaid on original video, results in the worst alignment. This outcome is anticipated, as the synthetic speech is generated using off the shelf text-to-speech (TTS) technology, which lacks an inherent understanding of temporal information across modalities. While the lip-sync scores of AV2AV model are close to ground truth as shown in \cite{choi2024av2av}, the results for the same model differs in ours as we only leverage speech generated via \textit{AV2AV-Speech} and skip the face synthesis. Consequently, this approach face the same problem as \textit{Synthetic} as there is no notion of lip-synchrony between the synthesized speech and original video thus resulting in sub-optimal alignment. This underscores the importance of introducing synchrony constraints in the training process.

Our approach fine-tunes the duration predictor by adding the lip-synchrony and duration losses as described in Section~\ref{ssec:dur_pred}. \textit{Ours} significantly outperform both \textit{Synthetic} and \textit{AV2AV-Speech} approaches in terms of lip-synchrony scores (p-value $<$ 0.05). This indicates that our methods are effective in enhancing the lip-synchrony of translated speech with the original video content.

While the motivation of our work is to improve lip-synchrony, it should \textbf{not} come at the cost of degraded naturalness in speech or translation quality~\cite{brannon2022dubbing}. Table~\ref{tab:donoharmresults} collects results for speech quality as measured by PESQ and translation quality as measured by BLASER-2.0 (reference-free metric) and ASR-BLEU (reference-based metric). We do not observe any degradation across four language pairs in either naturalness or translation quality.



\section{Ablations}
\label{sec:ablate}

\subsection{Duration Prediction}
\label{ssec:ablate_dur}


This ablation study assesses the impact of lip-sync loss (LS loss), duration loss (D. loss), and model initialization on lip-sync performance for English to Spanish translation, measured using LSE-C (higher is better) and LSE-D (lower is better). As shown in Table~\ref{tab:abla_dur}, our model (\textit{LS+D FT}), which incorporates both lip-sync and duration loss and is fine-tuned with parameters initialized from AV2AV pre-trained checkpoint, achieves the best performance with the highest LSE-C (2.45) and lowest LSE-D (10.68). \textit{LS+D} model, which is randomly initialized achieves worse results in comparison to ours (LSE-C: 1.96, LSE-D: 11.03). \textit{LS FT} model which is fine-tuned only with lip-sync loss without duration loss performs the worst with LSE-C of 1.76 and LSE-D of 11.70. These results indicate that combining both losses and leveraging pre-trained models significantly enhances lip-sync quality in translation tasks.


\begin{table}[t]
\begin{adjustbox}{max width=\columnwidth}
\begin{tabular}{l|c|c|c|c|c}
\textbf{System}    & \textbf{LS loss} & \textbf{D. loss} & \textbf{From Scratch?} & \textbf{LSE-C} & \textbf{LSE-D} \\ \hline
\textbf{LS+D FT}      & \greencheck & \greencheck & \redcross & 2.45  & 10.68 \\ \hline
\textbf{LS+D}       & \greencheck & \greencheck & \greencheck & 1.96  & 11.03 \\ \hline
\textbf{LS FT}      & \greencheck & \redcross & \redcross & 1.76  & 11.70  \\ 
\end{tabular}
\end{adjustbox}
\caption{Ablation study to understand the affect of lip-sync loss (LS loss), duration loss (D. loss) and initializing the model from pre-trained checkpoint for English to Spanish language direction.}
\label{tab:abla_dur}
\end{table}


\begin{figure}[t]
  \centering
  \includegraphics[width=.85\linewidth]{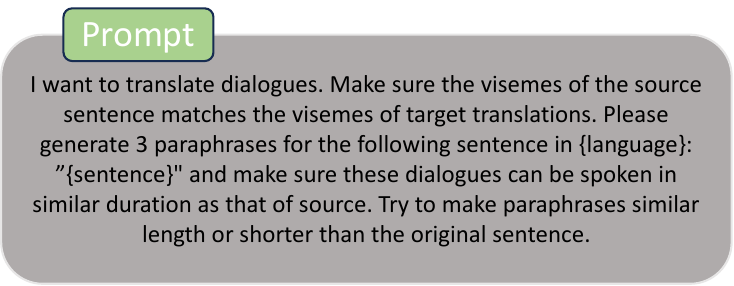}
   \caption{Prompt for generating paraphrases using Claude 3.0 Sonnet.}
   \label{fig:prompt}
   \vspace{-.2in}
\end{figure}
\vspace{-.1in}


\subsection{Unit-to-Unit Translation with Paraphrasings}
\label{ssec:u2u}


We sought to determine whether improved lip-synchrony could be achieved by using a unit-to-unit translation model that generates variations aligned with the original video. Training such a model requires a parallel corpus of translated speech. However, existing datasets, like LRS3, lack parallel source sentences with multiple target translations. To address this, we created multiple translation options from English (the source language) into our target languages (Spanish, Portuguese, Italian, and French). For each English sample, we first generated a target translation using Amazon Translate, which was then refined using a large language model (Claude 3.0 Sonnet) with a specific prompt in Figure~\ref{fig:prompt}. Finally, these translations were converted into speech using Amazon Polly.

We leveraged the pre-trained AV2AV encoder-decoder model \cite{choi2024av2av} specifically designed for unit-to-unit translation and fine-tuned it on the (spoken) paraphrased translations. During the fine-tuning process, we ensure that each batch is organized such that four targets (3 paraphrases + 1 original) corresponding to a source input are contained within the same batch. The unit-encoder receives a source language token $\langle L_s \rangle$, which indicates the language to be comprehended, along with the source AV speech units $\mathbf{u}_s = \{u^i_s\}_{i=1}^{T_s}$, where $T_s$ represents the number of units. The unit-decoder then takes a target language token $\langle L_t \rangle$, which determines the output language, and uses its previous predictions to autoregressively predict the next AV speech unit  in the target language. 



Although this new training approach of unit-to-unit translation model with paraphrases achieve better lip-synchrony scores over our strong system (Table~\ref{tab:paraphrase}), this comes at a cost to translation quality which violates our initial motivation of ``not trading off lip-synchrony improvements over speech translation quality and naturalness". We hypothesize that the drop in translation quality is due to LRS3 containing many short sentences (c.f. avg. duration per video in Table~\ref{tab:lrs3_stats}). While generating paraphrases, LLM changes the meaning of the sentence for e.g. ``And that's powerful'' is translated as ``Y eso es poderoso.'' (And that is powerful.) and paraphrased as ``Es muy fuerte.'' (It is very strong.) which is not a direct translation of source.

\begin{table}[t]
\begin{adjustbox}{max width=\columnwidth}
\begin{tabular}{l|c|c|c|c}
                  & \textbf{LSE-D ($\uparrow$)} & \textbf{LSE-C ($\downarrow$)} & \textbf{BLASER} & \textbf{ASR-BLEU} \\ \hline
\textbf{Ours}              &  10.68              &  2.45               & 0.797  & 35.18    \\ \hline
\textbf{Ours + Updated TM} &  10.18             &     2.91            & 0.766  & 32.18  \\  
\end{tabular}
\end{adjustbox}
\caption{Ablation study with the updated translation model (Ours + Updated TM) on translation variations. These results are on the English-Spanish language pair, but we observed similar trends across all four language pairs.}
\label{tab:paraphrase}
\end{table}

\begin{table}[t]
\begin{adjustbox}{width=0.75\columnwidth,center}
\centering
\renewcommand{\arraystretch}{1.5} 
\setlength{\tabcolsep}{10pt} 
\begin{tabular}{l|c|c}
\textbf{System} & \textbf{LSE-D $\uparrow$} & \textbf{LSE-C $\downarrow$} \\ \hline
\textbf{Ours + Updated TM} & 10.18 & 2.91 \\ \hline
\textbf{Length Match} & 10.35 & 2.66 \\ \hline
\textbf{Original Translation} & 10.67 & 2.60 \\ 
\end{tabular}
\end{adjustbox}
\caption{Comparison of \textit{Ours+Updated TM} with length matching and original translation approaches for the English-Spanish setting.}
\label{tab:2stage_abla}
\end{table}


To determine whether fine-tuning on translations that closely match the original speech length improves lip-synchrony, we selected the best length-matched paraphrase based on audio-visual (AV) units based on the principle of isometric translations producing better dubbing quality~\cite{lakew2021isometricmt}. Instead of training on four target translations, we trained the model using only this selected paraphrase. Surprisingly, the results worsened compared to our \textit{Ours+Updated TM} approach, as shown by the \textit{Length Match} results in Table~\ref{tab:2stage_abla} (LSE-D: 10.18 vs. 10.35). 
Additionally, when we used speech generated via text-to-speech (TTS) from the original translation (i.e., translation from Amazon Translate), we observed a significant degradation in LSE* metrics compared to \textit{Ours+Updated TM} -- specifically, LSE-D: 10.18 vs. 10.67.



\section{Conclusions}
\label{sec:conclusion}

This study focuses on generating speech that aligns seamlessly with the original video, avoiding the need for facial synthesis and ensuring high-quality dubbing. Our AVS2S framework incorporates lip-synchrony and duration loss to enhance the alignment between speech and lip movements in audio-visual translation models. By concentrating solely on improving lip-synchrony without altering facial features, our approach demonstrates significant improvements. While our approach is effective, further research is required to refine paraphrase generation, explore viseme variations across languages, and extend evaluations. We also aim to explore longer speeches, allowing for more nuanced paraphrasing that closely mirrors the original sentences.

\newpage
\vfill
\renewcommand{\normalsize}{\fontsize{9}{10}\selectfont}
\normalsize

\newpage



\bibliographystyle{IEEEbib}
\bibliography{strings,refs}

\end{document}